\documentclass{revtex4}
\usepackage{graphicx}
\usepackage{fancyhdr}
\usepackage{amsmath,amssymb}
\usepackage{color}

\begin{document}

\title{Grand Unified Theories and Higgs Physics}


\author{T. Yamashita}
\affiliation{School of Medicine, Aichi Medical University, Nagakute 480-1195, JAPAN}

\begin{abstract}
The grand unified theories are theoretically well motivated, but they typically 
 have less direct indications on the low energy physics and it is not easy to test them. 
Here, we discuss a scenario of them which naturally solves the so-called doublet-triplet 
 splitting problem and, at the same time, generally predicts characteristic collider 
 phenomenology. 
Then, we may get a hint on the breaking of the grand unified symmetry 
 at the on-going and next-generation collider experiments. 
\end{abstract}

\maketitle

\thispagestyle{fancy}


%%%%%%%%%%%%%%%%%%%%%%%%%%%%%%%%%%
\section{Introduction}
Since a resonance consistent with the standard model (SM) Higgs field was 
 reported~\cite{126GeVHiggs}, most people consider that the SM is now being confirmed, 
 at least as an effective theory valid below the TeV scale. 
Then, the next question that we ask is what will come as the physics beyond the SM. 
The reason we ask it is that the SM still has some problems and puzzles, such as the 
hierarchy problem and the charge quantization problem. 
Here, we emphasize that the latter requires a tuning at least as fine as $10^{-10}$ 
 to explain why the hydrogen atom is (almost) neutral. 
Thus, if we mind the former problem (as often happens), 
 the latter also should be taken care. 
A simple solution to the latter is to extend the gauge group to a semi-simple one. 
Looking back the history of the physics, which is that of the unification, 
 it is reasonable to take the idea of the grand unification~\cite{GUTs} seriously. 

Supposing the unification of the three forces of the SM, 
 the matter fields are also to be unified. 
This unification works perfectly for the SM fermions: the five multiplets in each 
 generation are unified into two in $SU(5)$ grand unified theories (GUTs). 
This is not trivial at all actually, as it becomes clear when people try to unify the 
 electroweak $SU(2)\times U(1)$ symmetry into $SU(3)$. 
Thus, this success strongly supports the idea of the grand unification. 
It is also to be commented that the idea can easily accommodate other ideas to solve 
 other SM problems: the supersymmetry (SUSY) for the hierarchy problem, 
 the conventional seesaw mechanism~\cite{seesaw} for the tiny 
 neutrino masses and the Leptogenesis~\cite{lepto} for the baryon asymmetry 
 of the universe.  
And if the $R$-parity is assumed, as usual in SUSY models, the 
 candidate dark matter is supplied. 
In addition, in the minimal model, the three running gauge couplings become almost 
 the same value at a superheavy scale, called the GUT scale: $2\times10^{16}{\rm GeV}$. 
This success of the gauge coupling unification (GCU) is so impressive that many people 
 tend to believe the SUSY-GUTs.
In this way, GUTs, especially SUSY-GUTs, potentially solve many of the problems/puzzles 
 in the SM and give an amazing by-product. 

On the other hand, the grand unification fails to unify the Yukawa interactions 
 and the Higgs field. 
The former is insisted as a consequence of the fermion unification and is not 
 necessarily bad for the third generations but not good for the lighter fermions. 
This issue is often called the wrong GUT relation and is to be taken care in model 
 building, while it is relatively easy to solve (see for example Ref.~\cite{nrMSSU(5)}). 
The latter requires a $SU(5)$ partner of the SM doublet Higgs fields. 
The minimal choice is to introduce a color-triplet partner to embed them into the 
 fundamental representation of the $SU(5)$ group. 
In SUSY models~\footnote{For the people who do not mind the fine tuning, the following 
 is not a problem neither.}, the triplet partner generates effective dimension five 
 operators that contribute to the nucleon decay~\cite{dim5PD}. 
In order to make the proton lifetime long enough without tuning, the triplet partner 
 should be much heavier than the GUT scale. 
It is not an easy task to realize naturally the mass splitting between such a 
 superheavy triplet and the weak scale doublet which originate from a common ($SU(5)$) 
 multiplet. 
This rather severe issue is called the doublet-triplet (DT) splitting problem and is 
 one of the biggest problem in the SUSY-GUTs. 
When we consider the grand unification seriously, these issues have to be dealt with. 

Since the idea of the grand unification is so attractive, this problem has been 
 attacked by many researchers for long time, and several solutions have been 
 proposed~\cite{SlidingSinglet, missingPARTN, missingVEV, pNG, Kawamura}.
They, however, all require some extension of the matter content, the grand unified 
 gauge group and/or spacetime geometry. 
These extensions bring rather large ambiguity on the gauge couplings around the GUT 
 scale due to the threshold corrections and so on. 
Then, what usually done are just to forget the ambiguity, to assume the corrections 
 are aligned not to affect the GCU or, at most, to make models so that the GCU is kept. 
At this stage, the GCU is no longer a success but just a constraint in model building. 
Here, however, we would like to stress that the success of the GCU is just a 
 by-product, and even without it the idea of the grand unification is attractive enough, 
 as mentioned above. 

Next, let us discuss indications of the grand unification on the low energy physics
 that we can detect. 
The most famous one is the nucleon decay. 
It is actually impressive prediction, but the information that we would get will be 
 rather little and thus it would be nice if there are some characteristic predictions 
 on the collider physics in addition. 
Unfortunately, since the GUT scale is so high, the decoupling 
 theorem~\cite{decoupling} makes it hopeless to detect the effects in most of the 
 SUSY-GUTs. 

Here, we would like to introduce a scenario of the SUSY-GUTs where the DT splitting 
 problem is naturally solved and an extraordinary collider phenomenology is generally 
 predicted~\cite{gGHU-DTS}.
Interestingly, this scenario, in a sense, can be regarded as an effective field 
 theoretical description of the SUSY-GUTs embedded into the heterotic string 
 theory~\cite{hetero} which could 
 treat the quantum gravity and explain the numbers of our spacetime~\cite{4d} and 
 of the generations~\cite{generations}. 
It is quite exiting if we can get some informations on the GUT breaking which might
 indicate the string theory at the on-going and next-generation collider experiments.

%%%%%%%%%%%%%%%%%%%%%%%%%%%%%%%%%%
\section{SUSY grand gauge-Higgs unification}

In this section we review the scenario proposed in Ref.~\cite{gGHU-DTS}. 
In the scenario, the Hosotani mechanism~\cite{Hosotani}, which works in higher 
 dimensional gauge theories, is applied to break the 
 GUT symmetry~\cite{gGHU}. 
In the Hosotani mechanism, the symmetry breaking occurs by the extradimensional 
 component of the gauge field which is a higher dimensional vector field. 
Thus, in this model, the Higgs field is unified with the gauge field, 
 and it is often called the gauge-Higgs unification especially when it is 
 applied to the electroweak symmetry breaking. 
In the present scenario, it is applied to the grand unified symmetry 
 breaking~\footnote{
Note that the Higgs field that is unified with the gauge field is an adjoint Higgs 
 field, in this case, and the SM doublet Higgs field is introduced as a matter field.}
 and named as grand gauge-Higgs unification~\cite{gGHU}.

An important point is that, in this case, the order parameter is not the 
 extradimensional component itself which is valued on the algebra, 
 but the Wilson loop which is valued on the group and thus free from the traceless 
 condition. 
Because of it, interestingly, a kind of the so-called missing VEV~\cite{missingVEV} 
 can be realized and the DT splitting problem is naturally solved 
 even in $SU(5)$ models~\cite{gGHU-DTS}. 
In this way, the application of the Hosotani mechanism to the GUT breaking in SUSY-GUTs 
 looks attractive. 

Naively thinking, the application to the GUT breaking seems reasonable since the 
 Higgs field that is unified with the gauge field behaves as an adjoint field. 
Actually, at the first stage of the study of this mechanism, it was applied to 
 the GUT breaking (or simpler toy model)~\cite{Hosotani,S1}. 
Unfortunately, however, chiral fermions can not be accommodated in these models 
 and thus these are phenomenologically less interesting. 
After the orbifold symmetry breaking~\cite{Kawamura} becomes famous among researchers 
 who works on phenomenological model building, 
 this mechanism have been applied mainly to the electroweak symmetry 
 breaking~\cite{S1Z2}.
This is because the orbifold symmetry breaking can extract fundamental-representational 
 components (with respect to the remaining subgroup of the original gauge group) 
 from the adjoint representation (with respect to the original group), 
 besides chiral fermions from the higher-dimensional fermions. 
Furthermore, in such models, the Higgs field are free from the quadratically divergent 
 radiative corrections to the mass term, thanks to the higher dimensional gauge 
 symmetry~\cite{hierarchy}.
In any case, now we know the orbifold symmetry breaking to realize chiral fermions, 
 and thus it is interesting to examine the application to the GUT breaking.
It might seem straightforward, but we immediately meet a difficulty.  
Namely, the adjoint scalar fields (with respect to the remaining gauge symmetry) 
 originated from the extra-dimensional components 
 tend to be projected out by the orbifold action 
 when chiral fermions are realized. 

This difficulty is shared with the heterotic string theory~\cite{hetero} 
 and, fortunately, a method, called diagonal embedding method~\cite{DiagonalEmbedding}, 
 to evade the difficulty is known. 
In Ref.~\cite{gGHU}, it is pointed out that the same method can be applied 
 in a field theoretical setup and thus we have an advantage that 
 it is much easier to calculate the quantum corrections that tell us 
 the positions of vacua. 
By this, the symmetry breaking pattern is controlled 
 by the dynamics described by the field theory irrelevantly to the ultraviolet theory, 
 in contrast to the orbifold breaking where it is chosen by hand. 
In addition, as mentioned above, the DT splitting problem, when the SUSY version is 
 considered, is naturally solved. 
Thus, this scenario is theoretically well motivated. 

Interestingly, this scenario generically gives particular predictions also 
 on the collider phenomenology. 
It is existence of light adjoint chiralmultiplets with masses of 
 the SUSY-breaking scale. 
The reason is as follows.
The adjoint Higgs field is a part of the gauge field and thus 
 massless at the tree level. 
Since the symmetry that ensures the masslessness is broken by the compactification, 
 the adjoint field gets mass corrections via the quantum corrections. 
As the radiative corrections would be vanishing when the SUSY was not broken, 
 the mass corrections are proportional to the SUSY breaking scale. 
In SUSY models, there are the SUSY partners of the adjoint scalar 
 which would be degenerate with the scalar if the SUSY is exact 
 and thus again has masses of at most the SUSY-breaking scale. 
Then, the whole the adjoint chiralmultiplet is predicted to be light 
 (if the SUSY-breaking scale is around the electroweak scale, as often expected) 
 while the components of $SU(5)/SU(3)\times SU(2)\times U(1)$ are eaten when 
 the unified $SU(5)$ gauge group is broken. 
Thus, color octet, weak triplet and singlet chiralmultiplets~\footnote{
In this scenario, models generally have a $\mathbb Z_2$ symmetry under which 
 these adjoint multiplets change the sign~\cite{gGHU-DTS}. 
Thus, this scenario can give a theoretical background of (SUSY) inert 
 models~\cite{inert}, though we consider a model where this $\mathbb Z_2$ symmetry 
 is broken.
}
 will appear in the 
 effective theory below the compactification scale (which is assumed to be around 
 the GUT scale), and they may be observed in the on-going and next-generation 
 collider experiments.

An immediate consequence of the adjoint chiralmultiplets is that the GCU is disturbed. 
As mentioned above, however, the GCU should be treated just as a constraint instead 
 of a success. 
It is easy to recover the GCU by adding further chiralmultiplets.
An example which is easily realized in this scenario and we consider here 
is two vectorlike pairs of $(\bf1,\bf2)_{-1/2}$, one of $(\bar \bf3,\bf1)_{-2/3}$ 
 and one of $(\bf1,\bf1)_{1}$, with which the GCU is realized at the GUT scale 
 and the unified gauge coupling is remains in the perturbative region: 
 $\alpha_G\sim0.3$.

Since the strong interaction is no longer asymptotically free (irrelevantly to the 
 choice of the additional fields to recover the GCU), 
 the QCD corrections are enhanced and thus the colored particles tend to be rather 
 heavy in this scenario. 
Although it is also interesting study to examine the extraordinary pattern of the 
 mass spectrum of the colored particles for the hadron colliders, 
 here we concentrate on the colorless fields: the singlet and the triplet. 
These additional fields couples to the two Higgs doublets of the minimal SUSY SM 
 (MSSM).
These couplings push the SM-like Higgs mass by the tree level $F$-term contribution 
 and thus the rather heavy Higgs mass around $125{\rm GeV}$ can be easily realized.
In addition, they cause mixing between the MSSM doublet Higgs fields and the adjoint 
 fields which result in modification of the coupling of the SM-like Higgs 
 fields~\cite{gGHU-pheno}. 
Such corrections may be measured at the linear collider. 
In the next section, we will discuss these issues in more detail.

%%%%%%%%%%%%%%%%%%%%%%%%%%%%%%%%%%
\section{phenomenology}

In order to examine the colorless sector, it is convenient to consider an effective 
 theory where the Higgs sector of the MSSM is extended with the singlet $S$ and 
 the triplet $\Delta$. 
The superpotential, in this case, is given as 
\begin{eqnarray}
W=\mu H_uH_d+\mu_\Delta{\rm tr}(\Delta^2)+\frac{\mu_S}2S^2
  +\lambda_\Delta H_u\Delta H_d + \lambda_SSH_uH_d, 
\end{eqnarray}
 where $H_u$ and $H_d$ are the MSSM doublet Higgs supermultiplets. 
Note that there are no self-couplings among $S$ and $\Delta$ although such couplings 
 are allowed by the symmetry at the level of the effective theory. 
This is because $S$ and $\Delta$ originate from the gauge field. 
Furthermore, this fact also insists the two additional couplings $\lambda_\Delta$ and 
 $\lambda_S$ to be related to the gauge couplings so that they are unified at 
 the GUT scale (with appropriate group theoretical factors).
Thus, this model is quite predictive (up to the dimensionful parameters).

For instance, taking the above example of the additional chiralmultiplets to recover 
 the GCU, the running of the gauge couplings are determined. 
The unified gauge coupling is used to fix the boundary values of 
 $\lambda_\Delta$ and $\lambda_S$ at the GUT scale, and we get 
\begin{equation}
 \lambda_\Delta=1.1,\qquad \lambda_S=0.26,
\end{equation}
 at the weak scale (within the 1-loop approximation). 
\begin{figure}[h]
\centering
\includegraphics[width=80mm]{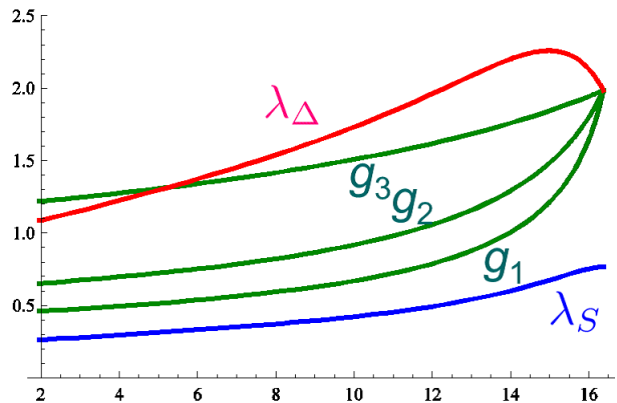}
\caption{An example of running couplings.}
\end{figure}
Using these predicted parameters, we can calculate the SM-like Higgs mass, the charged 
 Higgs mass in therms of the CP-odd Higgs mass, deviations of the SM-like Higgs 
 couplings from the corresponding SM values, and so on. 
Here, we just mention that the deviations are typically of order a few percents and 
 thus can be tested at the linear collider. 
The details of the results will be shown in Refs.~\cite{gGHU-pheno, Taniguchi}.

%%%%%%%%%%%%%%%%%%%%%%%%%%%%%%%%%%
\section{Summary}

In this article, we introduce the supersymmetric version~\cite{gGHU-DTS} of the grand
 gauge-Higgs unification scenario~\cite{gGHU} where the grand unified gauge symmetry 
 is broken by the Hosotani mechanism~\cite{Hosotani}. 
Interestingly, in this scenario, the doublet-triplet splitting problem can be solved 
 naturally even in $SU(5)$ models~\cite{gGHU-DTS}, thanks to the phase nature of the 
 Hosotani mechanism. 
In addition, it generally predicts the existence of light adjoint chiralmultiplets: 
 the color octet, the weak triplet and the neutral singlet. 
Their mass is around the supersymmetry-breaking scale, which is often assumed to be 
 the TeV scale, and thus there is a chance to detect them at the on-going and 
 next-generation collider experiments.

Due to the color octet chiralmultiplet, the QCD is no longer asymptotic free, and the 
 QCD corrections are typically enhanced. 
This suggests that the additional colored particles become rather heavy. 
Thus, we concentrate on the colorless fields~\cite{gGHU-pheno}, 
 though it is also an interesting work to examine the mass spectrum of the colored 
 particles. 
Then, the effective theory of this scenario becomes the one with an extended Higgs 
 sector: the neutral triplet and singlet are added. 
Since these are unified to the gauge field, they do not have self couplings and their 
 couplings are related to the unified gauge coupling. 
This fact makes the model very predictive. 
For instance, we can calculate the SM-like Higgs mass, the charged 
 Higgs mass in therms of the CP-odd Higgs mass, deviations of the SM-like Higgs 
 couplings from the corresponding SM values, and so on, up to the ambiguity due to 
 the dimensionful parameters. 
Although the details are referred to Refs.~\cite{gGHU-pheno, Taniguchi}, 
 we emphasize that the linear collider is expected 
 since the deviations are typically of order a few percents which are in its reach.

\begin{acknowledgments}
This article is partly based on a work in collaboration with M.~Kakizaki, S.~Kanemura 
 and H.~Taniguchi which is still in progress~\cite{gGHU-pheno}. 
\end{acknowledgments}

\bigskip % extra skip inserted

\end{document}